\documentclass[bibnotes,prl,twocolumn,superscriptaddress]{revtex4}%
\usepackage{epsfig,dsfont,amssymb,amsmath,amsthm,amsfonts,amsbsy,mathrsfs}
\usepackage{graphicx}
\usepackage{bm}
\begin{document}
\title{Two-photon dynamics in coherent Rydberg atomic ensemble}
\author{Bing He}
\affiliation{Institute for Quantum Science and Technology, University of Calgary, Calgary, Alberta T2N 1N4, Canada}
\affiliation{Department of Physics, University of Arkansas, Fayetteville, AR 72701, USA}
\author{A. V. Sharypov}
\affiliation{Kirensky Institute of Physics, 50 Akademgorodok, Krasnoyarsk, 660036, Russia}
\affiliation{Siberian Federal University, 79 Svobodny Av., Krasnoyarsk, 660041, Russia}
\author{Jiteng Sheng}
\affiliation{Department of Physics, University of Arkansas, Fayetteville, AR 72701, USA}
\author{Christoph Simon}
\affiliation{Institute for Quantum Science and Technology, University of Calgary, Calgary, Alberta T2N 1N4, Canada}
\author{Min Xiao}
\email{mxiao@uark.edu}
\affiliation{Department of Physics, University of Arkansas, Fayetteville, AR 72701, USA}
\affiliation{National Laboratory of Solid State Microstructures and School of Physics, Nanjing University, Nanjing 210093, China}

\begin{abstract}
We study the interaction of two photons in a Rydberg atomic ensemble
under the condition of electromagnetically induced transparency, combining 
a semi-classical approach for pulse
propagation and a complete quantum treatment for quantum state evolution.
We find that the blockade regime is not suitable for implementing
photon-photon cross-phase modulation due to pulse absorption and dispersion. However, approximately ideal cross-phase 
modulation can be realized based on relatively weak interactions, with
counter-propagating and transversely separated pulses.
\end{abstract}
\maketitle

\begin{figure}[b!]
\vspace{-0cm}
\centering
\epsfig{file=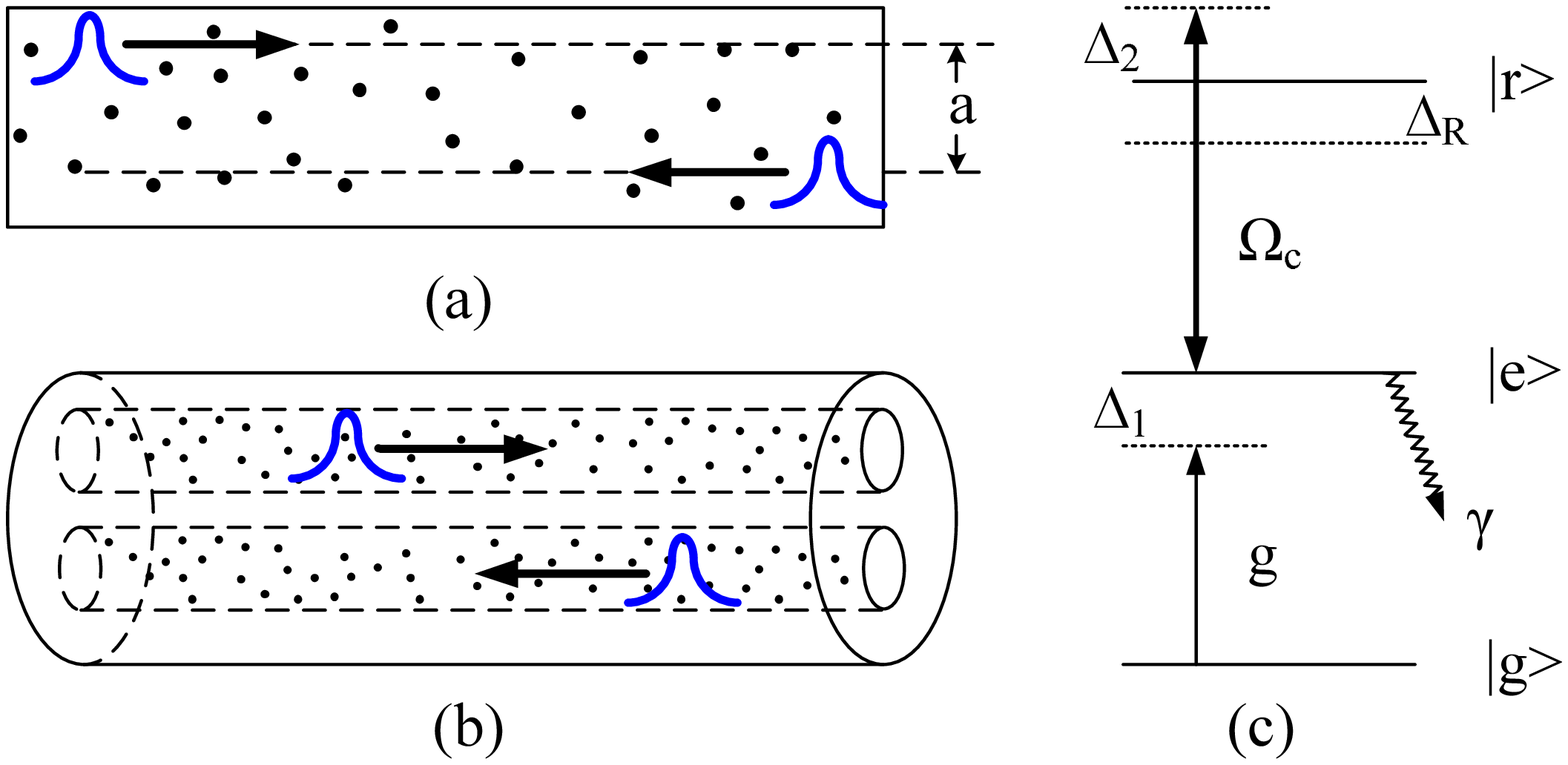,width=0.85\linewidth,clip=} 
{\vspace{-0.2cm}\caption{ (color online) (a) Two single photon pulses propagate in a Rydberg atomic ensemble. (b) Two pulses propagate in two parallel waveguides filled with Rydberg atoms. For the insignificant diffraction or the propagation in (b), the pulse profiles for the numerical simulations in Figs. 3 and 4 can be approximated as one-dimensional. (c) Atomic level scheme for the system. Without pulse interaction there is $\Delta_1+\Delta_2=0$ under the EIT condition. Here $\Delta_1=\omega_{eg}-\omega_p$, 
and $\Delta_2=\omega_{re}-\omega_c$ ($\omega_p$ is the input pulses' central frequency, and $\omega_c$ is the frequency of the pump beam). The Rydberg level is shifted 
by $\Delta_R$ due to the interaction with another pulse.   }}
\vspace{-0cm}
\end{figure}

Strong nonlinearity at the single-photon level is desirable to the realization of all-optical quantum devices. 
Ensembles of highly excited Rydberg atoms under electromagnetically induced transparency (EIT) condition combine the advantages of strong atom-field coupling without significant absorption and non-local atomic interaction, and have attracted intensive experimental \cite{EIT0,EIT1,EIT2,EIT3,EIT4,EIT5,EIT6} and theoretical studies \cite{EIT-0, EIT-1,EIT-2, EIT-3, EIT-4, EIT-5, EIT-6, EIT-7, EIT-8, EIT-9} recently. The strong correlation directly between single photons inside Rydberg atomic ensemble was observed \cite{EIT6}, and the formation of a Wigner crystal of individual photons is also predicted \cite{EIT-7}. When such interaction is applied to implement the cross-phase modulation (XPM) between two individual photons with a non-zero relative velocity as in Fig. 1 \cite{g1,g2,g3,g4}, a main difference from a single probe beam propagation in Rydberg EIT medium \cite{EIT0,EIT1,EIT2,EIT3,EIT4,EIT5, EIT6, EIT-0, EIT-1,EIT-2, EIT-3, EIT-4, EIT-5, EIT-6, EIT-7, EIT-8, EIT-9} is that no steady state exists for the pulses, because their interaction varies with the relative distance, pulse velocity that changes pulse sizes, as well as the absorption in medium. The realistic time-dependence in the inherent nonlinear dynamics makes a complete solution of the problem rather challenging. With the combination of a semi-classical approach for pulse propagation and a complete quantum approach for pulse quantum state evolution, we find a realistic picture for the dynamical process by showing the concerned figures of merit. We show that our proposed setup outperforms the previously considered
Rydberg blockade regime \cite{g3} clearly in terms of much lower photon absorption and negligible group velocity dispersion.

The detailed two-photon XPM via Rydberg EIT is as follows. One respectively couples the far-away input photons 
to cold Rydberg atoms under the EIT condition to form the light-matter quasi-particle called dark-state polariton (DSP) \cite{f-l}. The spatial distribution of the pulses necessitates a quantum many-body description of the process. The prepared DSPs are in the state 
$|1\rangle_l=\int d^3xf_l({\bf x})\hat{\Psi}^\dagger_l({\bf x})|0\rangle$ for $l=1, 2$, where $f_l({\bf x})$ 
are their snapshots with $\int d^3x|f_l({\bf x})|^2=1$, and $\hat{\Psi}({\bf x})=\cos\theta \hat{\mathcal{E}}({\bf x})-\sin\theta \hat{S}({\bf x})$ as the superposition of electromagnetic field operator $\hat{{\cal E}}({\bf x})$ 
and Rydberg spin-wave field operator $\hat{S}({\bf x})$ is the DSP field operator. The many-body version of the atom-field Hamiltonian 
\begin{eqnarray}
H^l_{AF}&=& -\frac{1}{2}\int d^3x \big \{\Omega_c\hat{S}_l^{\dagger}({\bf x})%
\hat{P}_l({\bf x})+g\sqrt{N}\hat{\mathcal{E}}_l^{\dagger}({\bf x})\hat{P}_l({\bf x})\nonumber\\
&+&H.c.\big\} +\int d^3x  \Delta_1\hat{P}^\dagger_l({\bf x})\hat{P}_l({\bf x})  \nonumber \\
&=&-\int d^3x \big\{\omega^+ \hat{\Phi}_{+,l}^\dagger\hat{\Phi}%
_{+,l}({\bf x})+\omega^- \hat{\Phi}_{-,l}^\dagger\hat{\Phi}_{-,l}({\bf x})\big\}~~~~
\label{cp}
\end{eqnarray}
($\hbar=1$) also involving the polarization field $\hat{P}({\bf x})$ for the excited level $|e\rangle$ can be diagonalized in terms of two bright-state polariton (BSP) fields
$\hat{\Phi}_+({\bf x})=\sin\theta \sin\phi \hat{\mathcal{E}}({\bf x})+\cos\phi \hat{P}
({\bf x})+\cos\theta\sin\phi \hat{S}({\bf x})$ and $\hat{\Phi}_-({\bf x})=\sin\theta \cos\phi \hat{\mathcal{E}}({\bf x})-\sin\phi \hat{P}
({\bf x})+\cos\theta\cos\phi \hat{S}({\bf x})$, where their spectrum $\omega^{\pm}=\frac{1}{2}(\Delta_1\pm\sqrt{\Delta_1^2+g^2N+\Omega_c^2})$ is a function of the input photon detuning $\Delta_1$, pump beam Rabi frequency $\Omega_c$ and atom density $N$. The combination coefficients for the polariton field operators satisfy the relations $\tan\theta=g\sqrt{N}/\Omega_c$ and $\tan 2\phi=\sqrt{g^2N+\Omega_c^2}/\Delta_1$ with $g$ as the atom-field coupling constant. When the DSPs get close 
to each other, the interaction 
\begin{eqnarray}
H_{I}&=&\int d^3x d^3x^{\prime}\hat{S}_1^{\dagger}({\bf x})\hat{S}%
_2^{\dagger}({\bf x}^{\prime})\Delta ({\bf x}-{\bf x}^{\prime})\hat{S}_2({\bf x}^{\prime})\hat{S}
_1({\bf x})~~~  \label{interaction}
\end{eqnarray}
between the pulses takes effect. Here we consider the Van der Waals (VdW) potential $\Delta({\bf x})=-C_6/|{\bf x}|^6$ in Rydberg atomic ensemble.  
Such interaction, however, also causes the transition of DSP to BSPs containing $\hat{P}_l({\bf x})$ components decaying at the rate $\gamma$. The decay of the $\hat{P}_l({\bf x})$ field is described by \cite{book} 
\begin{eqnarray}
H^l_{D}&=&i\sqrt{\gamma}\int d^3x \big\{\hat{P}_l({\bf x})
\hat{\xi}_l^{\dagger}({\bf x},t)-\hat{P}_l^{\dagger}({\bf x})\hat{\xi}_l({\bf x},t)\big\},
\label{diss}
\end{eqnarray}
with the white-noise operators of the reservoirs satisfying $[\hat{\xi}_l({\bf x},t),\hat{\xi}^{\dagger}_l({\bf x}^{\prime},t^{\prime})]=\delta^3({\bf x}-{\bf x}^{\prime})\delta(t-t^{\prime})$. The evolved pulse quantum state under all above mentioned factors should be close to the ideal output $e^{i\varphi}|1\rangle_1|1\rangle_2$ ($\varphi$ is a uniform one) for realizing a photon-photon XPM.

Before studying the input's quantum state evolution, one needs to ascertain the pulses' propagation in the medium, so that their interaction time should be known. The absorption and dispersion of the pulses can be found in a semi-classical approach \cite{EIT-95, book2} that treats the input pulses as the classical fields $E_{l}({\bf x})$, which are equivalent to the averages $\langle\hat{{\cal E}}_l({\bf x})\rangle$ of the quantum fields (up to a constant). In this framework the atom-field coupling is described by the following equations for the atomic density matrix elements \cite{EIT-95}:
\begin{subequations}
\begin{equation}
\dot{\rho}_{eg}=-(\frac{\gamma_{eg}}{2}+i\Delta_1)\rho_{eg}+i\frac{\mu_{eg}}{2}E_l(%
\rho_{ee}-\rho_{gg})-i\frac{\Omega_c}{2}\rho_{rg},  \label{a2}
\end{equation}
\begin{equation}
\dot{\rho}_{rg}=-\big(\frac{\gamma_{rg}}{2}+i(\Delta_1+\Delta_2+\Delta_R)\big)%
\rho_{rg}-i\frac{\Omega_c}{2}\rho_{eg}+i\frac{\mu_{eg}}{2}E_{l}\rho_{re},  \label{a3}
\end{equation}
\end{subequations}
where $\mu_{ij}$ are the transition dipole matrix elements and $\gamma_{ij}$
the decay rates of the relevant levels. The interaction with another pulse shifts the energy level of $|r\rangle$ and hence adds an extra term $\Delta_R({\bf x},t)=\sin^2 \theta T(t)\int d^3{\bf x}^{\prime} \Delta({\bf x}-{\bf x}^{\prime})\langle \hat{\Psi}_{3-l}^{\dagger}\hat{\Psi}_{3-l}\rangle({\bf x}^{\prime},t)$ 
to the detuning $\Delta_2$ of the pump beam, where $T(t)$ is the time-dependent transmission rate. This practice of reducing the interaction effect to a c-number detuning $\Delta_R$ is equivalent to a mean field treatment for 
the spin-wave fields in (\ref{interaction}). 
One has the time-dependent solution 
\begin{eqnarray}
\left(
\begin{array}
[c]{c}%
\rho_{eg}(t)\\
\rho_{rg}(t)
\end{array}
\right)&=&-i\frac{1}{2}\mu_{eg}\int_0^t d\tau e^{\int_\tau^t dt'\hat{M}(t')}
\left(
\begin{array}
[c]{c}%
E_{l}\\
0
\end{array}
\right)
\label{t-d}
\end{eqnarray}
to (\ref{a2})-(\ref{a3}) under the weak drive approximation \cite{EIT-95, book2} for single photons, where 
$$ \hat{M}(t)=\left(
\begin{array}
[c]{cc}
-(\frac{\gamma}{2}+i\Delta_1) & -i\frac{\Omega_c}{2}\\
-i\frac{\Omega_c}{2} & -\frac{\gamma_{rg}}{2}-i(\Delta_1+\Delta_2+\Delta_R(t))
\end{array}
\right)$$
with $\gamma_{eg}=\gamma$.
It is straightforward to obtain the time-dependent refractive index and decay rate from the susceptibility $\chi^{(1)}(t)=-2N\mu_{eg}\rho_{eg}(t)/(\epsilon_0 E_{l})$ based on (\ref{t-d}).

\begin{figure}[t!]
\vspace{0cm}
\centering
\epsfig{file=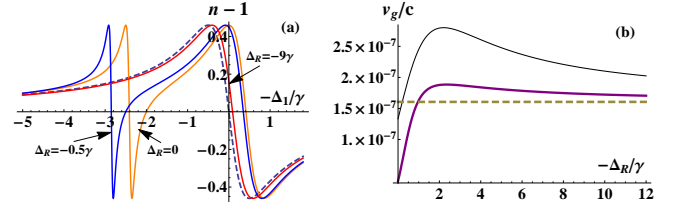,width=1.02\linewidth,clip=} 
{\vspace{-0.5cm}\caption{(color online) (a) Shift of refractive curves with increasing $\Delta_R< 0$. Here $n-1=1/2~\mbox{Re}\{\chi^{(1)}\}$ \cite{book2} is obtained from the approximated solution to (\ref{a2})-(\ref{a3}) in slow light regime, where $\Delta_R$ changes slightly during the decay time in the order of $1/\gamma$. We take the photon detuning $\Delta_1=2\gamma$ and the pump detuning $\Delta_2=-2\gamma$ under the EIT condition with $\Delta_R=0$. The pulses' initial group velocity under the EIT condition is set as $v_g=10$m/s ($v_g=c/n_g$ with $n_g=n+\omega_p\partial n/\partial \omega_p$), while the pump Rabi frequency is $\Omega_c=2\gamma$. The excited level $|e\rangle$ is $5P_{1/2}$ of $~^{87}$Rb. The dashed curve is that for the two level system of the corresponding parameters. The minus sign of the horizontal axes label comes from our definition of $\Delta_1$. (b) Group velocity $v_g$ vs $\Delta_R$ with the same pulse and pump detuning as in (a). The thicker solid curve is for $\Omega_c=2\gamma$, while the thinner is about $\Omega_c=4\gamma$. The dashed line is the group velocity of the corresponding two-level system.      }}
\vspace{-0cm}
\end{figure}

When two pulses approach each other, one phenomenon that could happen is known as Rydberg blockade. For the red-detuned photons ($\Delta_1>0$) the rising magnitude of negative $\Delta_R$ constantly shifts the refractive index curve going through the EIT point at a certain detuning $\Delta_1$ toward that of the corresponding two-level system. In the limit $|\Delta_R|\gg \gamma$ the system will virtually turn into a two-level one; see Fig. 2(a). One signature of Rydberg blockade is a platform of nearly unchanging group velocity shown in Fig. 2(b). In the blockade regime the pulse group velocity asymptotically tends to that of the corresponding two-level system; only those with $\Delta_1\leq 0.5\gamma$ in Fig. 2 can reach the speed of light $c$ with growing negative $\Delta_R$. 

The pulses will enter the superluminal regime characterized by anomalous dispersion \cite{book2}, which is accompanied by huge dissipation, if the interaction induced detuning $\Delta_R$ of the positive sign is gradually added to the pump beam of the system in Fig. 2. Equivalently this phenomenon happens to the blue-detuned single-photon pulses in the presence of the attractive VdW potential. This danger of completely damping the input photons should be avoided in practice.  

We therefore focus on red-detuned photons coupled to ensemble and propagating toward each other under the attractive interaction. As the pulses get closer, they will expand spatially because the characteristic size of their distributions $\langle \hat{\Psi}_l^{\dagger}\hat{\Psi}_l\rangle(z,t)$ is proportional to the average of the distributed group velocity $v_g(z,t)$ over the pulses. This modifies the $\Delta_R$ calculated with the relative distance and absorption of the pulses, which constantly keep changing as well. We use a numerical algorithm to simulate this dynamical process. From the coordinate origin $Z=0$ situated on the center of one pulse, the longitudinal relative distance $-L\leq Z\leq L$ to the other pulse's center throughout their motion is divided into $n_d$ grids. The detuning $\Delta_R$ at the $i$-th ($0<i\leq n_d-1$) position is calculated with the pulse size and transmission rate at the $i-1$-th position. Together with the obtained numerical values of $\Delta_R$ at the previous positions of $0 \leq k \leq i-1$, it is plugged into (\ref{t-d}) for the numerical integral to find the susceptibility $\chi^{(1)}$. In the same way the updated group velocity and transmission rate from the susceptibility at the position $i$ will be used to calculate the $\Delta_R$ at the $i+1$-th position. Running the iterative procedure with sufficiently large grid number $n_d$ approaches the real pulse motion.

\begin{figure}[t!]
\vspace{0cm}
\centering
\epsfig{file=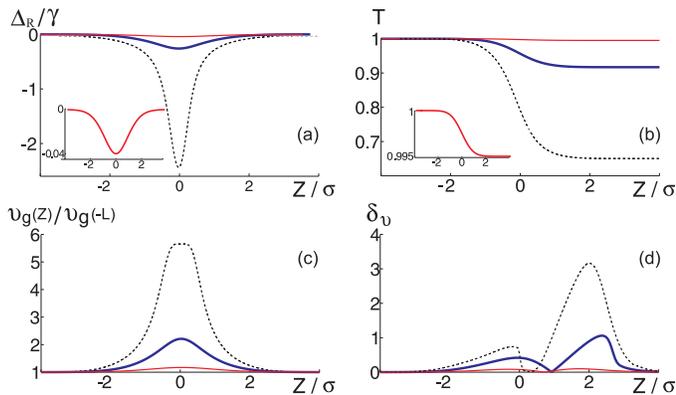,width=1.05\linewidth,clip=} 
{\vspace{-0.5cm}\caption{(color online) Numerical simulation for pulse motion. Here we adopt the relative distance coordinate $Z$ as a substitute for the time scale. We use $|g\rangle=5S_{1/2}$, $|e\rangle=5P_{1/2}$, and $|r\rangle=82D_{3/2}$ of $ ^{87}$Rb, with the VdW coefficient $|C_6|=8500~ \mbox{GHz}\cdot \mu \mbox{m}^6$ \cite{v-d} and $\gamma=2\pi \cdot 5.75~ \mbox{MHz}$. The system parameters 
are chosen as $Ng^2/\Omega_c^2=0.75 \cdot 10^7$ ($v_g(-L)=10$ m/s), $\Omega_c=2\gamma$, and $\Delta_1=-\Delta_2=2\gamma$.
The input Gaussian shaped pulses with $f(z)=\big(\frac{1}{\sqrt{\pi}\sigma}\big)^{\frac{1}{2}}e^{-\frac{1}{2}z^2/\sigma^2}$ have the initial size $\sigma=11.1~ \mu\mbox{m}$, with the corresponding bandwidth well fitted into the EIT window. The dashed curves are about the transverse separation $a=0.58~\sigma$, the thicker solid ones for $a=\sigma$, and the thinner solid ones for $a=1.5~\sigma$. The iterative step size for the numerical simulation is $0.005~\sigma$. (a) Interaction induced $\Delta_R(Z)$ at pulse centers. The insertion is the refined plot for $a=1.5~\sigma$. (b) Transmission rate $T(Z)=\exp(-k_p\int_{-L}^Z dy \mbox{Im} \{\chi^{(1)}(y)\})$. (c) Group velocity $v_g(Z)=c/n_g(Z)$
at pulse centers. The most transversely adjacent situation shows a velocity platform near Rydberg blockade. (d) Group velocity deviation ratio $\delta_v(Z)$. For the same $Z$, more extending pulses have higher $\delta_v$ due to more spatially inhomogeneous interaction.
}}
\vspace{-0cm}
\end{figure} 

Figure 3 illustrates an example of pulse motion found by the above mentioned numerical method. As 
shown in Figs. 3(a) and 3(b), the greater interaction between more transversely adjacent pulses 
is inseparable with the more significant pulse losses. In where Rydberg blockade starts to 
manifest, the accumulated pulse absorption has been harmful to the survival of the interacting photons (see Figs. 3(b) and 3(c)). The pulse absorption rate and group velocity in the blockade regime tend to those 
for a two-level system with the corresponding system parameters, so the only way to reduce the pulse loss 
in the blockade regime is using a higher photon detuning $\Delta_1$. However, one trade-off for doing so is to require a narrower pulse bandwidth (correspondingly a longer pulse size) to fit into the smaller EIT window, incurring a more prominent effect measured by the ratio $\delta_v(Z)=|\{v_g(Z,\sigma(Z))-v_g(Z,0)\}/v_g(Z,0)|$ shown in Fig. 3(d). Here $v_g(Z,\sigma(Z))$ is the group velocity at the location of the characteristic longitudinal size $\sigma(Z)$ from the pulse center, and $v_g(Z,0)$ is that at the pulse center. 
The non-uniform group velocity distribution over pulses (in the co-moving coordinate with the pulse centers) indicated by the ratio is equivalent to a group velocity dispersion that could make the pulses totally disappear even without absorption. Another disadvantage for large pulse size $\sigma$ is that the detuning value $\Delta_R$ from the spatially distributed pulses (proportional to $1/\sigma^6$ for the VdW potential) will be below the magnitude for a significant XPM. Our results thus show that
in the blockade regime considered in \cite{g3} the imperfections due to absorption and others are
actually much more problematic.

The next target is to understand the real-time evolution of the DSP state $|1\rangle_1|1\rangle_2$ given before (\ref{cp}). Under the perfect EIT condition, there is the approximation $\langle\hat{\sigma}_{gr}\rangle=-\mu_{eg}E_l/\Omega_c$ ($\hat{\sigma}_{gr}=|g\rangle\langle r|$) 
or its quantum many-body version $\hat{S}_l(z)= -(g\sqrt{N}/\Omega_c)\hat{{\cal E}}_l(z)$ after neglecting the non-adiabatic corrections for the narrow-band pulses, implying the identical propagation of the quantized DSP field with the electromagnetic field treated as classical in (\ref{a2})-(\ref{a3}) \cite{book2}. In the suitable 
weak interaction regime we find for the two-photon process, such as the most transversely separated pulses in Fig. 3 (corresponding to the refractive curves close to that of $\Delta_R=0$ in Fig. 2(a)), this approximation still holds with a small ratio $\Delta_R/\Omega_c$. The kinetic Hamiltonian for the slowly moving DSPs in the weak interaction regime can, therefore, be constructed as $H_K=-\sum_l iv_{g,l}(t)\int dz\hat{\Psi}_l^\dagger(z)\partial_z\hat{\Psi}_l(z)$, where the group velocity $v_{g,l}(t)$ is determined with (\ref{a2})-(\ref{a3}). Meanwhile, for a slow light with $\cos\theta\ll 1$, the BSPs interact very slightly with the DSPs and among themselves because they contain negligible Rydberg excitation. Their quick decoupling from the system and decaying into the environment allow one to treat the BSPs as motionless oscillations, though their group velocities can be read from their spectrum in (\ref{cp}). 

Our method for pulse state evolution is to adopt the joint evolution $U(t,0)$ as the time-ordered exponential $\mbox{T}e^{-i\int_0^t d\tau \{ H(\tau)+H_D(\tau)\}}$ on the initial state $|\psi_{in}\rangle=|1\rangle_1|1\rangle_2|0\rangle_c$ as the product of the input pulse state and the 
reservoir vacuum state $|0\rangle_c$. Tracing out the reservoir degrees of freedom in the evolved state $U(t,0)|\psi_{in}\rangle$ gives the evolved system state. 
We have three non-commutative items ($H_K$, $H_{AF}$ and $H_I$) in $H(t)$, as well as the dissipation Hamiltonian $H_D(t)$ of (\ref{diss}), for the joint evolution operator $U(t,0)$. Directly applying $U(t,0)$ on the DSP operators in $|\psi_{in}\rangle$ is impossible, as it is equivalent to analytically solving a nonlinear Langevin equation. One technique to circumvent the difficulty is the factorization of an evolution operator into the relatively tractable ones \cite{d}. For our problem we have $U(t,0)=U_K(t,0)U_{AF}(t,0)U_I(t,0)U_D(t,0)$ \cite{note}. Among the factorized processes $U_X(t,0)=\mbox{T}\exp\{-i\int_0^t d\tau \tilde{H}_X(\tau)\}$, for $X=K,AF,I$ and $D$, $\tilde{H}_K$ and $\tilde{H}_D$ are indifferent to 
their original form $H_K$ and $H_D$, respectively. The operator $U_D(t,0)$ takes no effect on $|\psi_{in}\rangle$, but the non-commutativity of $H_D$ with $H_{AF}$ makes the BSP field operators in $H_{AF}$ become those in $\tilde{H}_{AF}$ as follows:  
\begin{eqnarray}
\hat{\Phi}_{\pm,l}(z) &\rightarrow &\hat{\Xi}_{\pm,l}(z,\tau)=e^{-\phi_{\pm}^2 \gamma (t-\tau)/2}\hat{\Phi}_{\pm,l}(z)\nonumber\\
&\pm &\sqrt{\gamma}\phi_{\pm}\int_{\tau}^t dt^{\prime}e^{-\phi_{\pm}^2\gamma(t^{\prime}-\tau)/2}\hat{\xi}_l(z),
\label{BSP}
\end{eqnarray}
where $\phi_{+(-)}=\cos\phi (\sin\phi)$. A sufficiently large $\gamma$ approximates the commutator $[\hat{\Xi}_{\pm,l}(z,\tau_1),\hat{\Xi}^\dagger_{\pm,l}(z',\tau_2)]=e^{-\gamma \phi_{\pm}|\tau_1-\tau_2|}\delta(z-z')$ as 
vanishing for $\tau_1\neq \tau_2$. Under 
this approximation the BSP operators in $U_I(t,0)$ also take the forms in (\ref{BSP}), 
hence the evolved state 
$\hat{U}_I(t,0)|\psi_{in}\rangle$   
\begin{eqnarray}
&&\big\{\int dz_1 dz_2 f(z_1)f(z_2)e^{-ic_3^4\int_0^t d\tau \Delta(z_1^\tau-z_2^\tau )}
\hat{\Psi}^\dagger_1(z_1)\hat{\Psi}^\dagger_2(z_2) \notag \\
&-&ic_3^3 \sum_{l=1}^2\int_0^t d\tau\int dz_1 dz_2
f(z_1)f(z_2) e^{-ic_3^4\int_0^\tau dt^{\prime}\Delta
(z_1^{t^{\prime}}-z_2^{t^{\prime}})}  \notag \\
&\times & \Delta (z_1^\tau-z_2^\tau)
( c_1 \hat{\Xi}^\dagger_{+,{l}}+c_2
\hat{\Xi}^\dagger_{-,{l}})(z_1,\tau)\hat{\Psi}^\dagger_{3-l}(z_2)\big\}|0\rangle_t,~~~~
\label{output}
\end{eqnarray}
(unnormalized) to the first order of $\cos\theta$, where the notations $c_1=\cos\theta\sin\phi$, $c_2=\cos\theta\cos\phi $, $c_3=\sin\theta$, $z^\tau=z+\int_{0}^{\tau}d\tau ^{\prime }v_{g,l}(\tau ^{\prime })$, and $|0\rangle_t=|0\rangle|0\rangle_c$ are used 
to simplify the result. The detailed procedure for deriving the evolved state is given in \cite{note}. The succeeding operation $U_{AF}$ only affects the BSP components in (\ref{output}), while $U_K$ displaces the coordinate of $\hat{\Psi}^\dagger_l(z_l)$. 

\begin{figure}[t!]
\vspace{0cm}
\centering
\epsfig{file=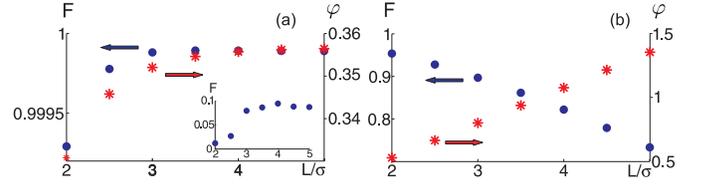,width=1.02\linewidth,clip=} 
{\vspace{-0.4cm}\caption{(color online) (a) Fidelity and cross phase of photon-photon XPM for two counter-propagating pulses with the transverse separation $a=1.5~\sigma$ in Fig. 3. $L$ is the medium size. The system parameters are the same as in Fig. 3. The insertion describes an imagined situation by reducing the initial pulse velocity to $10^{-2}$ m/s. (b) Fidelity and cross phase for two pulses propagating together along two tracks separated by 
$a=1.5~\sigma$. Due to pulse absorption, their group velocity is not stable in such co-propagation (for example, 
it drops from $11.007$ m/s to $11.002$ m/s from $L=2 \sigma$ to $5 \sigma$).  }}
\vspace{0cm}
\end{figure} 

The interaction potential $\Delta(z_1-z_2)$ renders the DSP part in (\ref{output}) no longer factorizable with respect to $z_1$ and $z_2$. This entangled piece deviates from the ideal 
output state $e^{i\varphi}|1\rangle_1|1\rangle_2$ with a uniform phase $\varphi$. We measure the degrees of such deviation by comparing the real 
output $|\psi_{out}\rangle=U(t,0)|\psi_{in}\rangle$
with a reference state $|\psi^0_{out}\rangle=U_K(t,0)U_{AF}(t,0)U_{D}(t,0)|\psi_{in}\rangle$. In the absence 
of $U_I(t,0)$ this reference keeps to be in the product 
state $|1\rangle_1|1\rangle_2|0\rangle_c$, even if the amplitude $f_l(z_l)$ in the output photon state $|1\rangle_l$ could be lowered due to any residual absorption. The output's fidelity $F$ with the ideal one and the associate cross phase $\varphi$ can thus be found from the overlap $\sqrt{F}e^{i\varphi}=\langle \psi^0_{out}|\psi_{out}\rangle$, where the two output states are normalized. 
Similar definitions for $F$ and $\varphi$ can be found in \cite{fid-1,fid-2}.

In Fig. 4 we plot the fidelity and cross phase for the most transversely separated pulses in Fig. 3. Due to the steep decay of the
VdW potential at long distances, both fidelity and cross phase for the counter-propagation in Fig. 4(a) quickly converge to fixed values with increasing medium size. A cross phase of $\pi$ rad that still 
keeps close to unit $F$ could be achieved if the VdW coefficient $|C_6|$, for example, is lifted by about nine times with a different Rydberg level. Contrary to a widely held notion, counter-propagation does not automatically ensure high fidelity; see \cite{fid-2}. The insertion of Fig. 4(a) shows the fidelity for an imagined motion of two pulses passing each other very slowly. The same propagation geometry indicates that the degrading fidelity in the slow motion comes from the growing pulse entanglement over a longer interaction time. In comparison we also study the co-propagating pulses in Fig. 4(b). The co-propagation exhibits considerable trade-off between $F$ and $\varphi$, and would be unfavorable for making large phases of good quality.

In summary, we have studied the process of two-photon interaction via a Rydberg atomic ensemble. Our approach based on the complete dynamics for both single atoms and ensemble enables a more realistic description of the situation without steady state. The previously considered regime near Rydberg blockade is found to be short of the favorable figures of merit for photon-photon XPM. We also prove that approximately ideal XPM creating considerable nonlinear phase can be realized with counter-propagating and transversely separated pulses that weakly interact with each other. The photon-photon XPM we have discussed can be the basis for an all-optical deterministic quantum phase gate.

B. H. and C. S. acknowledge the support by AITF and NSERC.
M.X. acknowledges the supports in part by NBRPC (Grant No. 2012CB921804) and NSFC (No. 11321063).
A. V. S. was supported by RFBR 12-02-31621. 

\begin{widetext}

\section*{Supplementary Information for ``Two-photon dynamics in coherent Rydberg atomic ensemble" }

\subsection*{A.~~~~~  Decomposition of Joint Evolution Operator}

\renewcommand{\theequation}{A-\arabic{equation}}
\setcounter{equation}{0}
The system Hamiltonian in the concerned problem consists of three parts. The atom-field coupling Hamiltonian for the ensemble is 
\begin{eqnarray}
H_{AF}&=& -\sum_{l=1}^2\int dz \big\{\omega^+ \hat{\Phi}_{+,l}^\dagger\hat{\Phi}
_{+,l}(z)+\omega^- \hat{\Phi}_{-,l}^\dagger\hat{\Phi}_{-,l}(z)\big\},
\label{couple}
\end{eqnarray}
where $\omega^{\pm}=\frac{1}{2}(\Delta_1\pm\sqrt{\Delta_1^2+g^2N+\Omega_c^2})$, and the bright-state polariton (BSP) fields are defined as
$$\hat{\Phi}_+(z)=\sin\theta \sin\phi \hat{\mathcal{E}}(z)+\cos\phi \hat{P}
(z)+\cos\theta\sin\phi \hat{S}(z),$$
$$\hat{\Phi}_-(z)=\sin\theta \cos\phi \hat{\mathcal{E}}({\bf x})-\sin\phi \hat{P}
(z)+\cos\theta\cos\phi \hat{S}(z).$$
 The polarization field $\hat{P}({\bf x})$ in the above is the continuous average $\sum_{i\in \Delta V}|g\rangle_i\langle e|/\sqrt{\Delta N}$ of the flip operators $|g\rangle_i\langle e|$ for the atoms inside a small volume $\Delta V$ around ${\bf x}$, which contains $\Delta N\gg 1$ atoms. So is the definition $\sum_{i\in \Delta V}|g\rangle_i\langle r|/\sqrt{\Delta N}$ for the spin-wave field $\hat{S}({\bf x})$.
The second part that describes the pulse interaction process is
\begin{eqnarray}
H_{I}&=&\int dz \int dz^{\prime}\hat{S}_1^{\dagger}(z)\hat{S}
_2^{\dagger}(z^{\prime})\Delta (z-z^{\prime})\hat{S}_2(z^{\prime})\hat{S}
_1(z)\nonumber\\
&=& \sum_{l=1}^2 \frac{1}{2} \int dz \big ( c_1 \hat{\Phi}^\dagger_{+,l}(z)+c_2 \hat{\Phi}^\dagger_{-,l}(z)+c_3 \hat{\Psi}^\dagger_l(z)
\big)\big ( c_1 \hat{\Phi}_{+,l}(z)+c_2 \hat{\Phi}_{-,l}(z)+c_3 \hat{\Psi}%
_l(z)\big)\hat{V}_l(z), 
\label{interaction}
\end{eqnarray}
where $\hat{V}_l(z)=\int dz^{\prime}\Delta(z-z^{\prime})\hat{\Lambda}^\dagger_{3-l}\hat{\Lambda}_{3-l}(z^{\prime},\tau)$ 
 with $\hat{\Lambda}_l=c_1 \hat{\Phi}_{+,l}+c_2 \hat{\Phi}_{-,l}+c_3 \hat{\Psi}_l$. 
Here we use the notations $c_1=\cos\theta\sin\phi$, $c_2=\cos\theta\cos\phi $, and $c_3=\sin\theta$ from the main text. The third part is the kinetic Hamilton 
\begin{eqnarray}
H_{K}&=&-i\sum_{l=1}^2 v_{g,l}(t)\int dz\hat{\Psi}_l^\dagger(z)\partial_z\hat{\Psi}_l(z),
\label{kinetic}
\end{eqnarray}
for the DSPs, where $v_{g,l}(t)$ is found with a semi-classical treatment of the atom-field coupling in the main text. Similarly the BSP kinetic Hamiltonian can be constructed with their group velocities $1/2c(1\pm \Delta_1/\sqrt{\Delta_1^2+g^2N+\Omega_c^2})$ from the spectrum in (\ref{couple}). In a slow light regime considered in the main text, the BSPs go much faster than and interact very slightly with the DSPs, while they decay into the environment. Such quick decoupling of the BSPs from the system allows one to approximate them as motionless oscillations, and this simplifies the coordinates for the BSP field operators in most equations below.  
In addition, the coupling between the polarization fields $\hat{P}_l$ and reservoir that leads to the dissipation is described by
\begin{eqnarray}
H_{D}&=&i\sum_{l=1}^2\sqrt{\gamma}\int dz \big\{\hat{P}_l(z)
\hat{\xi}_l^{\dagger}(z,t)-\hat{P}_l^{\dagger}(z)\hat{\xi}_l(z,t)\big\}\nonumber\\
&=& i\sum_l \sqrt{\gamma}\int dz \big\{\cos \phi~ \hat{\Phi}_{+,l}(z)\hat{\xi}_l^{\dagger}(z,\tau)
-\sin \phi~ \hat{\Phi}_{-,l}(z)\hat{\xi}_l^{\dagger}(z,\tau)-H.c.\big\},
\label{diss}
\end{eqnarray}
where the random-variable noise operators satisfy $[\hat{\xi}_l(z,t),\hat{\xi}^{\dagger}_l(z^{\prime},t^{\prime})]=\delta(z-z^{\prime})\delta(t-t^{\prime})$. The infinitesimal action 
of the joint evolution $U(t,0)=\mbox {T}e^{-i\int_0^t d\tau (H(\tau)+H_D(\tau)}$, where $H=H_K+H_{AF}+H_I$, 
on the field operators and the joint quantum state of the system and reservoir gives rise to the exact Langevin equation about the system operators and the exact master equation about the system state, respectively \cite{book}. The solution to these equations are difficult to find in the presence of the nonlinear term in (\ref{interaction}).

Here we present a different approach to find the transformation 
$U(t,0)\hat{\Psi}_l(z)U^{\dagger}(t,0)$ by factorizing the joint evolution operator $U(t,0)$ into relatively tractable processes. First, we separate the kinetic part out of the total evolution operator 
as follows: 
\begin{eqnarray}
\mbox {T}\exp\{-i\int_0^t d\tau H(\tau)\}&=&\mbox {T}\exp\{-i\int_0^t
d\tau H_K(\tau)\}~ \mbox {T}\exp\big \{-i\int_0^t d\tau U^{\dagger}_K (\tau,0) \big(H_{AF}+H_I+H_D\big)U_K(\tau,0)\big\},  
\label{1}
\end{eqnarray}
where $U_K(\tau,0)= \mbox {T}\exp\{-i\int_0^\tau d t^{\prime}H_K(t^{\prime})\}$. The proof for this exact factorization can be found in \cite{d}. 
The interaction Hamiltonian in the second time-ordered exponential of the above becomes
\begin{eqnarray}
&&H_I(\tau)=U^{\dagger}_K (\tau,0)H_I(\tau)U_K(\tau,0)  \notag \\
&= & \frac{1}{2}\sum_{l=1}^2\int dz \big ( c_1 \hat{\Phi}^\dagger_{+,l}(z)+c_2 \hat{\Phi}^\dagger_{-,l}(z)
+c_3 \hat{\Psi}^\dagger_l(z^{\tau,l})\big) \big ( c_1 \hat{\Phi}_{+,l}(z)+c_2 \hat{\Phi}_{-,l}(z)+c_3 \hat{\Psi}_l(z^{\tau,l})\big)\hat{V}_l(z,\tau),  
\end{eqnarray}
where 
\begin{eqnarray}
\hat{V}_l(z,\tau)&=&\int d\zeta \Delta(z-\zeta) \big ( c_1 \hat{\Phi}^\dagger_{+,3-l}(\zeta)+c_2 \hat{\Phi}^\dagger_{-,3-l}(\zeta)+c_3 \hat{\Psi}^\dagger_{3-l}(\zeta^{\tau,3-l})\big) \big ( c_1 \hat{\Phi}_{+,3-l}(\zeta)+c_2 
\hat{\Phi}_{-,3-l}(\zeta)+c_3 \hat{\Psi}_{3-l}(\zeta^{\tau,3-l})\big)\nonumber\\
\end{eqnarray}
with $z^{\tau,l}=z-\int_0^\tau v_{g,l}(t^{\prime}) dt^{\prime}$. The effect of the above transformation is the displacement of the coordinates for the DSP field operators. The other terms in the second time-ordered exponential of (\ref{1}) are not changed. 

Secondly, the system-reservoir coupling process in the second time-ordered exponential of (\ref{1}) is separated out to the right side as follows:
\begin{eqnarray}
&&\mbox {T}\exp\{-i\int_0^t d\tau \big(H_{AF}+H_I(\tau)+H_D\big)\}\nonumber\\
&=&\mbox {T}\exp\{-i\int_0^t
d\tau U_D(t,\tau)\big(H_{AF}+H_I(\tau)\big)U_D^\dagger(t,\tau)\}~ 
\mbox {T}\exp\big \{-i\int_0^t d\tau H_D(\tau)\big\},  
\label{2}
\end{eqnarray} 
where $U_D (t,\tau)=\mbox{T} \exp\{-i\int_\tau^t dt' \hat{H}_{D}(t')\}$. 
In the first time-ordered exponential of the above, the BSP fields will be transformed to 
\begin{eqnarray}
&&U_D (t,\tau)\hat{\Phi}_{+,l}(z)U^\dagger_D(t,\tau)\nonumber\\
&=& e^{-\cos^2\phi \gamma (t-\tau)/2}\hat{\Phi}_{+,l}(z)+
\underbrace{\cos\phi\sqrt{\gamma}\int_{\tau}^t dt^{\prime}
e^{-\cos^2\phi\gamma(t^{\prime}-\tau)/2}\hat{\xi}_l(z,t^{\prime})}_{\hat{n}_{+,l}(z,\tau)} 
\equiv \hat{\Xi}_{+,l}(z,\tau); 
\label{b1}
\end{eqnarray}
\begin{eqnarray}
&&U_D (t,\tau)\hat{\Phi}_{-,l}(z)U^\dagger_D(t,\tau)\nonumber \\
&=& e^{-\sin^2\phi \gamma (t-\tau)/2}\hat{\Phi}_{-,l}(z)-
\underbrace{\sin\phi\sqrt{\gamma}\int_\tau^t dt^{\prime}e^{-\sin^2\phi
\gamma(t^{\prime}-\tau)/2}\hat{\xi}_l(z,t^{\prime})}_{\hat{n}_{-,l}(z,\tau)} \equiv \hat{\Xi}_{-,l}(z,\tau).  
\label{b2}
\end{eqnarray}
The transformed BSP operators therefore satisfy the following
commutation relations: 
\begin{eqnarray}  
&&[\hat{\Xi}_{+,l}(z,\tau_1), \hat{\Xi}^\dagger_{+,l}(z^{\prime},\tau_2)]=
e^{-\cos^2\phi \gamma|\tau_1-\tau_2|/2}\delta(z-z^{\prime})\equiv g_1(\tau_1,\tau_2)\delta(z-z^{\prime}); \nonumber\\ 
&&[\hat{\Xi}_{-,l}(z,\tau_1), \hat{\Xi}^\dagger_{-,l}(z^{\prime},\tau_2)]= 
e^{-\sin^2\phi \gamma|\tau_1-\tau_2|/2}\delta(z-z^{\prime})\equiv g_2(\tau_1,\tau_2)\delta(z-z^{\prime}).  
\label{cc}
\end{eqnarray}
Then the first time-ordered exponential in (\ref{2}) takes the form 
\begin{eqnarray}
&&\mbox{T}e^{-i\frac{1}{2}\sum_{l=1}^2\int_0^t d\tau\int dz \hat{\Pi}^\dagger_l(z,\tau)\hat{\Pi}_l(z,\tau)\hat{V}_l(z,\tau)
+ i\sum_l \int_0^t d\tau\int dz \{\omega^+ \hat{\Xi}_{+,l}^\dagger\hat{\Xi}_{+,l}(z,\tau)+\omega^- \hat{\Xi}_{-,l}^\dagger\hat{\Xi}_{-,j}(z,\tau)\}}
\label{mide}
\end{eqnarray}
with $\hat{\Pi}_l(z,\tau)= c_1 \hat{\Xi}_{+,l}(z)+c_2 \hat{\Xi}_{-,l}(z)+c_3 \hat{\Psi}_l(z^{\tau,l})$.

The action of the second term inside the time-ordered exponential in (\ref{mide}) can be further separated out 
as in Eq. (\ref{1}),
and the accompanying effect is to transform the BSP operators inside the other time-ordered exponential as follows:
\begin{eqnarray}
&&\mbox{T}e^{-i\int_0^\tau dt^{\prime}\sum_l \int d\zeta \{\omega^+ \hat{\Xi}_{+,l}^\dagger\hat{\Xi}_{+,l}
(\zeta)+\omega^- \hat{\Xi}_{-,j}^\dagger\hat{\Xi}
_{-,l}(\zeta)\}}~ \hat{\Xi}_{+,l}(z, \tau) 
~\mbox{T}e^{i\int_0^\tau dt^{\prime}\sum_l \int d\zeta \{\omega^+ \hat{\Xi}_{+,l}^\dagger\hat{\Xi}_{+,l}
(\zeta)+\omega^- \hat{\Xi}_{-,l}^\dagger\hat{\Xi}
_{-,l}(\zeta)\}} \notag \\
&=& \hat{\Xi}_{+,l}(z,\tau)+i\int_0^\tau dt_1\omega^+ g_1(t_1,0)\hat{\Xi}_{+,l}(z,t_1)+\frac{(-1)}{2!}\int_0^\tau dt_2\int_0^\tau dt_1
(\omega^+)^2 g_1(t_2,0)g_1(t_1,0)\hat{\Xi}_{+,l}(z,t_2)+\cdots\nonumber\\
&\equiv & \hat{\Xi}'_{+,l}(z,\tau);\nonumber\\
&&\mbox{T}e^{-i\int_0^\tau dt^{\prime}\sum_l \int d\zeta \{\omega^+ \hat{\Xi}_{+,l}^\dagger\hat{\Xi}_{+,l}
(\zeta)+\omega^- \hat{\Xi}_{-,l}^\dagger\hat{\Xi}
_{-,l}(\zeta)\}}~ \hat{\Xi}_{+,l}(z) 
~\mbox{T}e^{i\int_0^\tau dt^{\prime}\sum_l \int d\zeta \{\omega^+ \hat{\Xi}_{+,l}^\dagger\hat{\Xi}_{+,l}
(\zeta)+\omega^- \hat{\Xi}_{-,l}^\dagger\hat{\Xi}
_{-,l}(\zeta)\}} \notag \\
&=& \hat{\Xi}_{-,l}(z,\tau)+i\int_0^\tau dt_1\omega^- g_2(t_1,0)\hat{\Xi}_{-,l}(z,t_1)+\frac{(-1)}{2!}\int_0^\tau dt_2\int_0^\tau dt_1
(\omega^-)^2 g_2(t_2,0)g_2(t_1,0)\hat{\Xi}_{-,l}(z,t_2)+\cdots\nonumber\\
&\equiv & \hat{\Xi}'_{-,l}(z,\tau).
\label{os}
\end{eqnarray}
Here we have neglected the mixing of the two transformed BSP fields due to their coupling to the same reservoir.
The pulse interaction process after the factorization now takes the form 
\begin{eqnarray}
U_{I}(t,0)&=&\mbox{T}\exp\big\{-i\frac{1}{2}\sum_{l=1}^2\int_0^t
d\tau\int dz \hat{\Theta}^\dagger_l(z,\tau)\hat{\Theta}_l(z,\tau)\hat{W}_l(z,\tau)\big\},
\end{eqnarray}
with $\hat{W}_l(z,\tau)=\int dz^{\prime}\Delta(z-z^{\prime})\hat{\Theta}^\dagger_{3-l}(z^{\prime},\tau)\hat{\Theta}_{3-l}(z^{\prime},\tau)$
and $\hat{\Theta}_l(z,\tau)=c_1 \hat{\Xi}'_{+,{l}}(z,\tau)+c_2 \hat{\Xi}'_{-,{l}}
(z,\tau)+c_3 \hat{\Psi}_{l}(z^{\tau,l})$. For a sufficiently large damping rate $\gamma$, the commutators in (\ref{cc})
can be regarded as vanishing for $\tau_1\neq \tau_2$, and then the BSP field operators in the above equation can be approximated as $\hat{\Xi}_{\pm,l}$ with $g_{1(2)}(t_1,t_2)\approx 0$ for any pair of $t_1$ and $t_2$. 

So far the joint evolution operator has been decomposed as 
\begin{eqnarray}
&&\mbox {T} e^{-i\int_0^t d\tau H(\tau)}  
=\mbox {T}e^{-i\int_0^t d\tau H_K(\tau)}~ 
\mbox{T} e^{i\int_0^t dt^{\prime}\sum_l\int d\zeta \{\omega^+ \hat{\Xi}%
_{+,l}^\dagger\hat{\Xi}_{+,l}(\zeta)+\omega^- \hat{\Xi}_{-,l}^\dagger\hat{\Xi}_{-,l}(\zeta)\}}  \notag \\
&\times &\mbox{T}e^{-i\frac{1}{2}\sum_{l=1}^2\int_0^t d\tau\int dz 
\hat{\Theta}^\dagger_l(z,\tau)\hat{\Theta}_l(z,\tau)\hat{W}_l(z,\tau)}~  
\mbox{T} e^{-i\int_0^t d\tau H_{D}(\tau)}  \notag \\
&\equiv & U_K(t,0)U_{AF}(t,0)U_I(t,0)U_D(t,0).
\label{decomposed}
\end{eqnarray}

\subsection*{B.~~~~~~Evolution of Pulse Quantum State} 
 \renewcommand{\theequation}{B-\arabic{equation}}
 \setcounter{equation}{0}
 
 Now we study the evolution of the joint state  
\begin{eqnarray}
|\psi_{in}\rangle=\int dz_1 f(z_1)\hat{\Psi}^{\dagger}_1(z_1)\int dz_2
f(z_2)\hat{\Psi}^{\dagger}_2(z_2)|0\rangle\otimes  |0\rangle_c,  \label{input2}
\end{eqnarray}
for two identical pulses, where $|0\rangle_c$ is the reservoir vacuum state, under $U(t,0)$. It is equivalent to finding the transformation $U(t,0)\hat{\Psi}^{\dagger}_l(z_l)U^\dagger(t,0)$ (or $U(t,0)\hat{\Psi}_l(z_l)U^\dagger(t,0)$) because $U(t,0)|0\rangle\otimes  |0\rangle_c=|0\rangle\otimes  |0\rangle_c\equiv |0\rangle_t$, i.e.
\begin{eqnarray}
U(t,0)|\psi_{in}\rangle&=&U(t,0)\int dz_1 f(z_1)\hat{\Psi}^{\dagger}_1(z_1)\int dz_2
f(z_2)\hat{\Psi}^{\dagger}_2(z_2)|0\rangle_t\nonumber\\
&=&\int dz_1 f(z_1) U(t,0)\hat{\Psi}^{\dagger}_1(z_1) U^\dagger(t,0)\int dz_2
f(z_2) U(t,0)\hat{\Psi}^{\dagger}_2(z_2) U^\dagger(t,0)|0\rangle_t.
\label{tt}
\end{eqnarray}
We will apply the decomposed form in (\ref{decomposed}) for the purpose.

The operation $U_D(t,0)$ does not change $|\psi_{in}\rangle$. 
The transformation by $\hat{U}_{I}(t,0)$ is found through 
\begin{eqnarray}
&&U_{I}(t,0)\hat{\Psi}_{l}(z_{l})U_{I}^{\dagger }(t,0)  
=\big(\mbox{T}e^{-ic_{3}^{2}\int_{0}^{t}d\tau \{U_{I}(t,\tau )\hat{W}_{l}(z_{l}^{-\tau},\tau )U_{I}^{\dagger }(t,\tau )\}}\big)^{\dagger }\hat{\Psi}_{l}(z_{l}) \nonumber \\
&+&ic_{3}\int_{0}^{t}d\tau \big(\mbox{T}e^{-ic_{3}^{2}\int_{0}^{\tau
}dt^{\prime }\{U_{I}(t,t^{\prime })\hat{W}_{l}(z_{l}^{-t^{\prime }},t^{\prime })
U_{I}^{\dagger }(t,t^{\prime })\}}\big)^{\dagger } 
\nonumber \\
&\times &U_{I}(t,\tau )\big (c_{1}\hat{\Xi}'
_{+,l}(z_{l}^{-\tau })+c_{2}\hat{\Xi}'
_{-,l}(z_{l}^{-\tau })\big )\hat{W}_{l}(z_{l}^{-\tau},\tau )
U_{I}^{\dagger }(t,\tau ),  
\label{exact}
\end{eqnarray}
where $z_{l}^{-\tau }=z_{l}+\int_{0}^{\tau }v_{g,l}(t^{\prime })dt^{\prime }$. 
This is an exact form obtained by expressing $\hat{U}_{I}(t,0)$ as an infinite
product of the small elements around each moment, which transform the DSP operator as follows:
\begin{eqnarray}
&&U_I(t+dt, t)\hat{\Psi}_{l}(z_{l})U^\dagger_I(t+dt, t)
=\big(\hat{I}+ic_{3}^{2}\hat{W}_{l}(z_{l}+\int_{0}^{t}dt^{\prime }v_{g,l}(t^{\prime }),t )dt\big)\hat{\Psi}_{l}(z_{l})\nonumber\\
&+&ic_3\big (c_{1}\hat{\Xi}'
_{+,l}(z_{l}^{-t })+c_{2}\hat{\Xi}'_{-,l}(z_{l}^{-t })\big )\hat{W}_{l}(z_{l}+\int_{0}^{t}d\tau ^{\prime }v_{g,l}(\tau ^{\prime }),t )dt.
\end{eqnarray}
The operation by $\hat{U}_{I}(t,\tau )$ inside the time-ordered exponential and integral of (\ref{exact}) 
can be further performed to obtain a form of this exact transform in terms of an infinite series.
There is the following commutator
\begin{eqnarray}
&&[\hat{W}_l(z_1,\tau),\hat{W}_l(z_2,\tau^{\prime})]=[\int
dz^{\prime}\Delta(z_1-z^{\prime})\hat{\Theta}^\dagger_{3-l}(z^{\prime},\tau)%
\hat{\Theta}_{3-l}(z^{\prime},\tau),\int
d\zeta^{\prime}\Delta(z_2-\zeta^{\prime})\hat{\Theta}^\dagger_{3-l}(\zeta^{%
\prime},\tau^{\prime})\hat{\Theta}_{3-l}(\zeta^{\prime},\tau^{\prime})] 
\notag \\
&\approx &\{ic_1^2e^{-\gamma
\sin^2\phi|\tau-\tau^{\prime}|/2}\sin g_1(\tau,\tau^{\prime})+
ic_2^2e^{-\gamma \cos^2\phi|\tau-\tau^{\prime}|/2}\sin
g_2(\tau,\tau^{\prime})\}\int dz^{\prime}\Delta(z_1-z^{\prime})\Delta(z_2-z^{\prime})\hat{\Theta}^\dagger_{3-l}(z^{\prime},\tau)\hat{%
\Theta}_{3-l}(z^{\prime},\tau^{\prime})\nonumber\\
\label{w-commutator}
\end{eqnarray}
for $\hat{W}_l(z,\tau)$. Together with the fact $e^{-\gamma \sin^2\phi|\tau-\tau^{\prime}|/2}\ll 1$, $e^{-\gamma
\cos^2\phi|\tau-\tau^{\prime}|/2}\ll 1$ for a sufficiently large damping rate $\gamma$ 
(this means a negligible correlation time window for the colored noises $\hat{n}_{\pm,l}$ introduced in Eqs. (\ref{b1}) and (\ref{b2})), the above commutator can be approximated as vanishing 
for $\tau\neq \tau'$ in a slow light regime with $|c_1|\ll 1$ and $|c_2|\ll 1$, which is created for the input photons under the EIT condition. Meanwhile one has $\sin g_{1(2)}(\tau,\tau^{\prime})=0$ for $\tau=\tau^{\prime}$.
Then there is the relation $\hat{U}_I(t,\tau)\hat{W}
_l(z_l^{-\tau},\tau)\hat{U}^\dagger_I(t,\tau)\approx \hat{W}_l(z_l^{-\tau},\tau)$ from the approximation of the vanishing commutator in (\ref{w-commutator}), and the non-Abelian phases in (\ref{exact}) can be
reduced to the Abelian ones due to such approximate commutativity of $\hat{W}
_l(z_l^{-t},t)$ at the different time. Moreover, in a slow light regime where the BSPs containing negligible Rydberg excitation quickly decouple from the system through decaying to the environment and escaping from the medium, the DSP components transformed back from the BSP components through the transformation $U_{I}(t,\tau) \big (c_1 \hat{\Xi}'_{+,l}(z_l)+c_2
\hat{\Xi}'_{-,l}(z_l) \big)U^\dagger_{I}(t,\tau)$ in the second term of (\ref{exact}) is negligible. 
Therefore, the DSP operator transformation in (\ref{exact}) can be finally approximated as
\begin{eqnarray}
U_I(t,0)\hat{\Psi}_l(z_l)U^\dagger_{I}(t,0) 
&=&  e^{ic_3^2\int_0^t d\tau \hat{W}_l(z_l^{-\tau})}\hat{\Psi}_l(z_l)+ic_3
\int_0^t d\tau e^{ ic_3^2\int_0^\tau dt^{\prime}\hat{W}_l(z_l^{-t^{\prime}})}  \notag
\\
&\times & \big ( c_1 \hat{\Xi}'_{+,l}(z^{-\tau}_l)+c_2
\hat{\Xi}'_{-,l}(z^{-\tau}_l)\big )\hat{W}_l
\big(z_l^{-\tau}\big)  \label{app}
\end{eqnarray}
in the regime considered in the main text. In Eq. (8) of the main text we express this DSP operator 
evolution with a further approximated form, considering the vanishing commutators in (\ref{cc}) for different time due to a sufficiently large damping rate $\gamma$.

To find the evolution for the state $|\psi_{in}\rangle$, one needs the following operation  
\begin{eqnarray}
&& U_I(t,0)\hat{\Psi}^\dagger_1(z_1)\hat{\Psi}^\dagger_2(z_2)|0\rangle_t=
\big \{e^{-ic_3^2\int_0^t d\tau \hat{W}_1(z_1^{-\tau})}\hat{\Psi}^\dagger_1(z_1)  
-ic_3 \int_0^t d\tau e^{ -ic_3^2\int_0^\tau dt^{\prime} \hat{W}_1(z_1^{-t^{\prime}})}\nonumber\\
&\times & \big ( c_1\hat{\Xi}'^\dagger_{+,1}(z^{-\tau}_1)+c_2
\hat{\Xi}'^\dagger_{-,1}(z^{-\tau}_1)\big ) \hat{W}_1\big(z_1^{-\tau}\big)\big\}  
\hat{\Psi}^\dagger_2(z_2)|0\rangle_t   \label{t}
\end{eqnarray}
based on (\ref{app}), where the relation $\hat{W}_2\big(z_2^{-\tau}\big)|0\rangle_t=0$ has been considered.
Similar to Eq. (\ref{app}), the phase operator $e^{-ic_3^2\int_0^t d\tau \hat{W}_1(z_1^{-\tau})}$ in the first term acts on the second DSP field operator $\hat{\Psi}^\dagger_2(z_2)$ via the transformation
\begin{eqnarray}
&&e^{-ic_3^2\int_0^t d\tau \hat{W}_1(z_1+\int_0^\tau
dt^{\prime}v_{g,1}(t^{\prime}))}\hat{\Psi}^\dagger_2(z_2)e^{ic_3^2\int_0^t
d\tau \hat{W}_1(z_1+\int_0^\tau dt^{\prime}v_{g,1}(t^{\prime}))}  
=e^{-ic_3^4\int_0^t d\tau \Delta \big (z_1-z_2+\int_{0}^{\tau}\{
v_{g,1}(t^{\prime\prime})-v_{g,2}(t^{\prime\prime})\} dt^{\prime\prime}\big )%
}\hat{\Psi}^\dagger_2(z_2)  \notag \\
&-&ic_3^3\int_0^t d\tau e^{-ic_3^4\int_0^\tau dt^{\prime}\Delta
\big(z_1-z_2+\int_{0}^{t^{\prime}}\{
v_{g,1}(t^{\prime\prime})-v_{g,2}(t^{\prime\prime})\} dt^{\prime\prime}\big )%
}\Delta \big (z_1-z_2+\int_{0}^{\tau}\{
v_{g,1}(t^{\prime\prime})-v_{g,2}(t^{\prime\prime})\} dt^{\prime\prime}) 
\notag \\
&\times & \big ( c_1 \hat{\Xi}'^\dagger_{+,{2}%
}(z^{-\tau}_2)+c_2 \hat{\Xi}'^\dagger_{-,{2}}(z^{-\tau}_2)\big).
\end{eqnarray}
The second term's action on $\hat{\Psi}^\dagger_2(z_2)$ is found through the commutator
\begin{eqnarray}
&&[ e^{ -ic_3^2\int_0^\tau dt^{\prime}\hat{W}_1(z_1+\int_0^{t^{\prime}}
d\tau^{\prime}v_{g,1}(\tau^{\prime}))} \hat{W}_1(z_1+\int_0^{\tau}
d\tau^{\prime}v_{g,1}(\tau^{\prime})),\hat{\Psi}_2^\dagger(z_2)]  \notag \\
&=&c_3^2 e^{-ic_3^4\int_0^\tau d\tau^{\prime}\Delta \big (%
z_1-z_2+\int_{0}^{\tau^{\prime}}\{
v_{g,1}(t^{\prime\prime})-v_{g,2}(t^{\prime\prime})\} dt^{\prime\prime}\big )%
}\Delta \big (z_1-z_2+\int_{0}^{\tau}\{
v_{g,1}(t^{\prime\prime})-v_{g,2}(t^{\prime\prime})\}\hat{\Psi}%
^\dagger_2(z_2)  \notag \\
&+& c_3 e^{-ic_3^4\int_0^{\tau} dt^{\prime}\Delta \big (z_1-z_2+%
\int_{0}^{t^{\prime}}\{
v_{g,1}(t^{\prime\prime})-v_{g,2}(t^{\prime\prime})\} dt^{\prime\prime}\big )%
}  \Delta \big (z_1-z_2+\int_{0}^{\tau}\{
v_{g,1}(t^{\prime\prime})-v_{g,2}(t^{\prime\prime})\} dt^{\prime\prime}\big )\nonumber\\
&\times & \big ( c_1 \hat{\Xi}'^\dagger_{+,{2}}(z^{-\tau}_2)+c_2
 \hat{\Xi}'^\dagger_{-,{2}}(z^{-\tau}_2)\big).
\end{eqnarray}
Putting all these together one will obtain the entangled state (unnormalized and to the first order of $c_{1(2)}$) 
\begin{eqnarray}
&& U_I(t,0)\int dz_1f(z_1)\hat{\Psi}%
^\dagger_1(z_1)\int dz_2f(z_2)\hat{\Psi}^\dagger_2(z_2)|0\rangle_t  \notag \\
&=& \int dz_1\int dz_2f(z_1)f(z_2)e^{-ic_3^4\int_0^t d\tau \Delta \big (%
z_1-z_2+\int_{0}^{\tau}\{
v_{g,1}(t^{\prime\prime})-v_{g,2}(t^{\prime\prime})\} dt^{\prime\prime}\big )%
}\hat{\Psi}^\dagger_1(z_1)\hat{\Psi}^\dagger_2(z_2)|0\rangle_t  \notag \\
&-&ic_3^3 \sum_{l=1}^2\int dz_1\int dz_2
f(z_1)f(z_2)\int_0^t d\tau e^{-ic_3^4\int_0^\tau dt^{\prime}\Delta
(z_1-z_2+\int_{0}^{t^{\prime}}\{v_{g,1}(t^{\prime\prime})-v_{g,2}(t^{\prime\prime})\})}  \notag \\
&\times & \Delta \big (z_1-z_2+\int_{0}^{\tau}\{
v_{g,1}(t^{\prime\prime})-v_{g,2}(t^{\prime\prime})\} dt^{\prime\prime}\big )%
\big ( c_1 \hat{\Xi}'^\dagger_{+,{3-l}}(z^{-\tau}_2)+c_2
\hat{\Xi}'^\dagger_{-,{3-l}}(z^{-\tau}_2)\big)\hat{\Psi}^\dagger_l(z_l)|0\rangle_t\nonumber\\ 
\label{o-state}
\end{eqnarray}
due to the evolution under $U_I(t,0)$.
After finding the above $U_I(t,0)|\psi_{in}\rangle$, it will be
straightforward to do the further transformations under $U_{AF}(t,0)$, which
transforms the BSP components as in (\ref{os}), and by $U_K(t,0)$, which displaces the DSP coordinates.
Tracing out the reservoir degrees of freedom makes no difference to the DSP part for the output state of
the system.

As we explain in the main text, the cross phase for the output state $|\psi_{out}\rangle=U(t,0)|\psi_{in}\rangle$ and its fidelity with an ideal output state from XPM are found through its overlap with the reference state $|\psi^0_{out}\rangle=U_K(t,0)U_{AF}(t,0)U_{D}(t,0)|\psi_{in}\rangle$. In the absence of the pulse interaction process, there is no BSP components in the reference state $|\psi^0_{out}\rangle$. The approximations we make for
the simplification of the evolved state in the main text, therefore, do not affect the values of the cross phase and fidelity in Fig. 4 of the main text.

\end{widetext}

\end{document}